\documentclass[aps,prl, twocolumn, showpacs, superscriptaddress]{revtex4-1}
\usepackage[utf8]{inputenc}
\usepackage{graphicx}
\usepackage{pstricks}
\usepackage{array}
\usepackage{natbib}
\usepackage{amsmath}
\usepackage{pifont}
\usepackage{dsfont}
\usepackage{multirow}
\usepackage{amssymb}
\usepackage{wasysym}

\def\be{\begin{equation}}
\def\ee{\end{equation}}
\def\tb{\textbf}

\begin{document} 

\title{AC Josephson effect without superconductivity}

\newcommand{\spsmsA}{Univ. Grenoble Alpes, INAC-SPSMS, F-38000 Grenoble, France}
\newcommand{\spsmsB}{CEA, INAC-SPSMS, F-38000 Grenoble, France}
\author{Benoit Gaury}
\affiliation{\spsmsA}
\affiliation{\spsmsB}
\author{Joseph Weston}
\affiliation{\spsmsA}
\affiliation{\spsmsB}
\author{Xavier Waintal}
\affiliation{\spsmsA}
\affiliation{\spsmsB}
\date{\today}

\begin{abstract}
Superconductivity derives its most salient features from the coherence of its
macroscopic wave function. The associated  physical phenomena have now moved
from exotic subjects to fundamental building blocks for quantum circuits such as
qubits or single photonic modes. Here, we theoretically find that the AC
Josephson effect---which transforms a DC voltage $V_b$ into an {\it oscillating}
signal $cos(2eV_b t/ \hbar)$---has a mesoscopic counterpart in normal
conductors. We show that on applying a DC voltage $V_b$ to an electronic
interferometer, there exists a universal transient regime where the current
oscillates at frequency $eV_b/h$.  This  effect is not limited by a
superconducting gap and could, in principle, be used to produce tunable AC
signals in the elusive $0.1-10$ THz ``terahertz gap''.
\end{abstract}

\maketitle

Superconductivity, a macroscopic quantum state, is described by a wave function
whose phase is physically significant. Indeed, quantum mechanical interference
effects are ubiquitous in superconducting systems, similar to those observed
at the microscopic scale in atomic physics. Since the eighties, such effects have also been
also observed at the mesoscopic (or nano) scale in condensed matter.
Most of the peculiar effects observed in superconductors  have an analogue
in ``normal'' quantum nanoelectronics: the DC SQUID (superconducting quantum
interference device) corresponds to the Aharonov-Bohm
effect~\cite{Batelaan2009}, supercurrents (at the origin of the Meissner effect)
correspond to the so-called
persistent currents~\cite{Imry2002,Levy1990,Harris2013}.  The AC Josephson
effect in superconductors is perhaps the most striking manifestation of these interference
effects at a macroscopic scale~\cite{Josephson1962}; a DC voltage bias $V_b$ applied across a
weak link between two superconductors creates an oscillating current with
frequency $2eV_b/h$. This voltage to frequency conversion is  used in metrology
to define the Volt in terms of the Second~\cite{Parker1967}, as well as in a
wealth of superconducting devices (RF-SQUIDs, quantum
bits)~\cite{Devoret2013}. Its origin is rather straightforward. The energy of
the left superconductor is $eV_b$ higher than the right one, so that its wave
function gets an extra oscillating factor $e^{-i 2e V_b t/\hbar}$. The junction
produces an interference between these two wave functions, hence the
oscillations. The absence of a corresponding mesoscopic effect in normal
conductors is in itself surprising from a theoretical perspective;
superconductivity is well described by the Bogoliubov-De Gennes
equation~\cite{degennesbook}, a simple extension of the
Schrödinger equation.
Here we report on this missing effect. We find that an oscillating signal $\cos
(eV_b t/\hbar)$ is generated in the transient regime that follows an abrupt
change of the bias voltage applied to a normal conductor.

Our story begins with a---two path---electronic Mach-Zehnder interferometer,
\begin{figure}[h!]
    \center
    \includegraphics[width=0.48\textwidth]{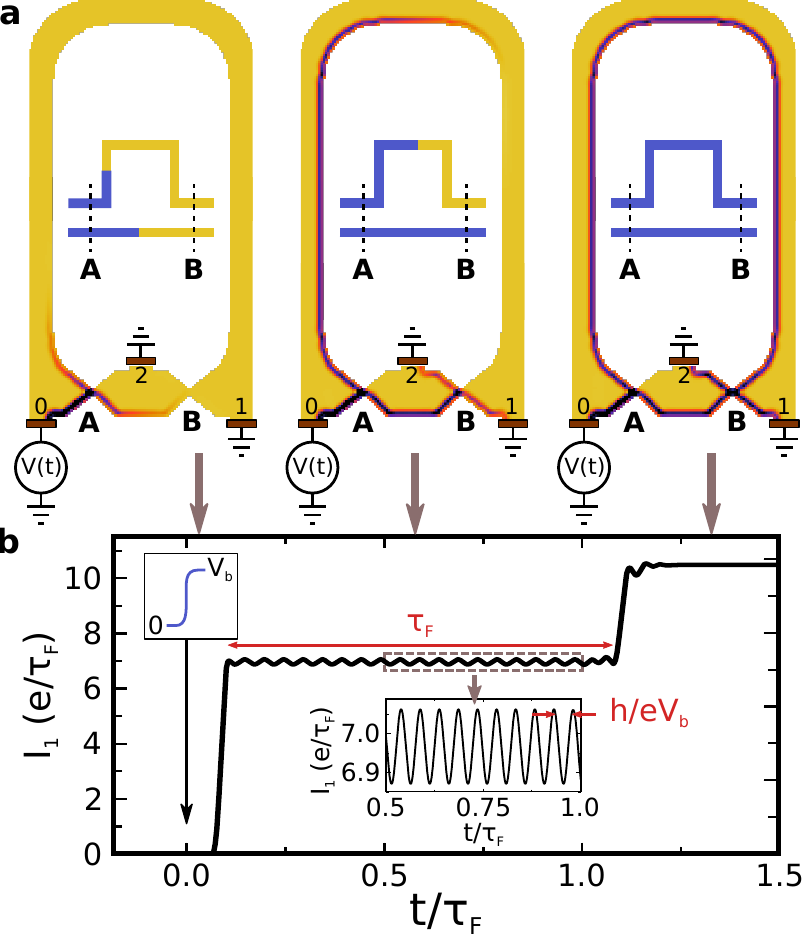}
    \caption{\label{mz} \tb{Raising the DC voltage bias on a Mach-Zehnder
    interferometer in the quantum Hall regime.} (\textbf{a}) Colour plot of the local
    electronic charge density (measured from equilibrium) in a 3 terminal
    Mach-Zehnder interferometer in the quantum Hall regime: from vanishing
    density (yellow), to $10^{11} cm^{-2}$ (red). At $t=0$, the voltage bias is
    raised from $V(t<0)=0$ to $V_b=20h/(e\tau_F)$. The three colour plots correspond to three
    snapshots for different times as indicated by the arrows.  A
    two-dimensional electron gas (yellow) is connected to three electrodes, the
    semi-transparent quantum point contacts A and B act as beamsplitters.
    Insets: schematics of the propagation of the voltage bias along the two
    arms of the interferometer. (\textbf{b})
    Transmitted current at contact 1. Upper
    inset: schematic of the raising of the bias voltage. Lower inset: zoom on
    the oscillations of the current.}
\end{figure}
sketched in Fig.~\ref{mz}a. This device, implemented in a two-dimensional gas
under high magnetic field, has lately become a rather standard tool of the
mesoscopic physicist~\cite{Heiblum_MachZender,20mum_20mK}. In the quantum Hall
regime the bulk of the electronic gas is insulating and the electronic
propagation only occurs on the edges of the sample. One can realize electronic
beam splitters with quantum point contacts, and in this way ensure that only two
paths are available for any travelling electrons. The sample is very asymmetric,
the upper arm being much longer than the lower one, which implies an extra time
of flight $\tau_F=L/v_g$ (with $L$ the extra length of the upper arm with
respect to the lower one and $v_g$ the group velocity of the edge state). At $t=0$
one raises the bias voltage applied on contact $0$ from $V(t<0)=0$ to
$V(t>\tau_P)=V_b$. While the exact manner in which the voltage is raised is
unimportant, the rise time $\tau_P$ must be sufficiently fast ($\tau_P<\tau_F$),
and the voltage drop spatially sharp enough (compared to
$L$)~\cite{Twave_formalism,fabryperot}. Fig.~\ref{mz}b shows the transmitted
current $I_1(t)$ as a function of time $t$, and we can discern three distinct
regimes.  In the beginning (Fig.~\ref{mz}a left) the voltage pulse did not have
enough time to propagate up to contact $1$, and $I_1(t)=0$. During a transient
regime of duration $\tau_F$ (Fig.~\ref{mz}a middle), the pulse has arrived at
contact $1$ from the lower arm but not yet from the upper one.  The current
increases to a
finite value. Finally (Fig.~\ref{mz}a right), the pulse arrives from the upper arm
and the current increases to its stationary value. 
The most noteworthy feature of Fig.~\ref{mz}b lies in the transient regime;
the current \emph{oscillates} with frequency $eV_b/h$ around 
a DC component.
This transient oscillatory regime is the mesoscopic analogue of the AC
Josephson effect. It is to the AC Josephson effect what persistent currents~\cite{Imry2002} are
to supercurrents.

The theory required to obtain this transient oscillatory regime is fundamentally simple.
Within the time-dependent scattering approach~\cite{Twave_formalism}, one finds that the wave function close to contact $0$ is a plane wave that acquires
an additional phase when the bias voltage is raised,
\be
\label{eq:pwave_phase}
\Psi_0(x,t)= \frac{1}{\sqrt{k}}e^{ikx -iEt/\hbar -ieV_b t \theta(t - x/v_g)/\hbar}
\ee
where $\theta(x)$ is the Heaviside function, $E$ is the incident energy of the
electron, $k$ the corresponding momentum, and the curved coordinate $x$ follows the
edge of the sample. We have assumed for simplicity a linear dispersion relation
$E(k)=\hbar v_g k$ and the condition $\tau_P\ll \tau_F$. We see from Eq.~(\ref{eq:pwave_phase})
that raising the voltage induces an oscillating phase difference $e^{ieV_b t/\hbar}$ between the front and the rear of the wave. One can consider
this phase difference as the time-dependent extension of the stationary case that was discussed
in~\cite{Levitov1996,fabryperot}. The device uses the delay time $\tau_F$ between
the two arms to create an interference between the rear and the front of the
wave function, generating the oscillatory behavior. In the transient
regime, the wave function close to contact $1$ is the superposition of the
contributions from the two paths and one finds,
\be
 \Psi_1(x,t)= \frac{e^{ikx -iEt/\hbar}}{\sqrt{k}} d_1(t,E)
\ee
with the total time-dependent transmission amplitude $d_1(t,E)$ given by,
\be
d_1(t,E)= d_u(E)e^{-iE\tau_F/\hbar}+d_l(E+eV_b) e^{-ieV_b t/\hbar}
\ee
The amplitudes $d_{u/l}$ for the upper/lower arm are given in terms of the transmission probabilities
$T_{A/B}$ of the quantum point contacts, $d_u=\sqrt{T_AT_B}$ and
$d_l=\sqrt{(1-T_A)(1-T_B)}$. Using the time-dependent generalization of the
Landauer formula~\cite{Twave_formalism}, 
\be
\label{landauer}
I_1(t)=(e/h) \int dE  |d_1(t,E)|^2 f(E)
\ee 
[$f(E)$ is the Fermi function, Eq.~(\ref{landauer}) includes the equilibrium
current injected from contact 0 which needs to be subtracted], we finally get
the current at contact 1 during the transient regime,
\begin{align}
\label{eq:boss}
I_{1}(t) =&\frac{e^2V_b}{h}  T_A T_B \nonumber\\
&+ \frac{e}{\pi\tau_F}\sqrt{T_A T_B (1-T_A)(1-T_B)}\cos\left(\frac{eV_bt}{\hbar} + \phi\right)
\end{align}
Eq.~(\ref{eq:boss}) is the main result of this letter and agrees with the
direct microscopic numerical calculations presented above. While the precise
coefficients depend on the particular interferometer considered, its structure
is totally general. It contains a DC term plus an AC term at frequency
$eV_b/h$; the amplitude of the AC current is of the order of $e/\tau_F$. For a
typical micrometer sized Mach-Zehnder interferometer, the amplitude of the AC
current is of the order of a few nA. 

The Mach-Zehnder interferometer is  simple conceptually, but
challenging experimentally in terms of the lithography (for the central
electrode $2$), low temperature and high magnetic field.
\begin{figure}[h!]
    \center
    \includegraphics[width=0.48\textwidth]{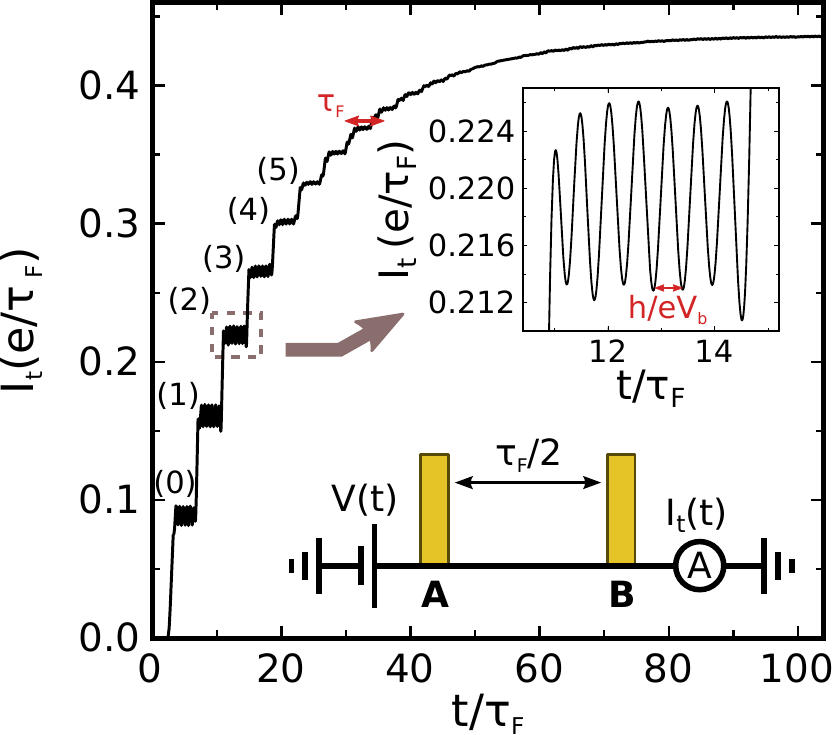}
    \caption{\label{fpcurrent}\tb{Raising the DC voltage bias on a Fabry-Perot
    interferometer.} Transmitted current  (in units
    of $e/\tau_F$, where $\tau_F$ is twice time of flight between the two
    barriers) as a function of time for a Fabry-Perot cavity. At $t=0$, the
    voltage bias is raised from $V(t<0)=0$ to $V_b=6h/(e\tau_F)$.  Upper inset: zoom on the
    oscillations of the current on a plateau.  Lower inset: schematic of the
    Fabry-Perot cavity ($T_a=T_b=0.1$). }
\end{figure}
Fabry-Perot cavities, in contrast, are ubiquitous and occur every time two barriers are put
in series. Examples include carbon nanotubes~\cite{FabryPerot_carbon_nano}, quantum Hall
systems~\cite{FabryPerot_pioneer} and
semi-conducting nanowires~\cite{InAsFP}. Fig.~\ref{fpcurrent} shows a sketch of the
Fabry-Perot geometry together with a numerical calculation of  the measured
current as a function of time. The $I_t(t)$ curve now features many steps that
correspond to the arrival of the path with direct transmission (0), the path
with one reflection on \emph{B} and \emph{A} (1), two reflections (2) and so on.
Again, each of these steps is accompanied by oscillations at the frequency
$eV_b/h$.  On decreasing the transparencies of the barriers, $T_A$ and
$T_B$, the Fabry-Perot resonances gradually become  true bound states and the
duration of the transient regime increases accordingly. This situation is very
close, mathematically, to the true Andreev bound states that occur in
a Josephson junction~\cite{Beenakker1991}.

Experimentally, progress in ultrafast quantum transport has been steady with
the recent demonstration of single electron sources resolved in
energy~\cite{Single_e_source} or time~\cite{Glattli2013} as well as measurements in the THz
range~\cite{McEuen2008}. For an easier observation of the transient AC signal, it
might be convenient to replace the abrupt voltage change described above by a
train of square pulses (with a ``slow'' period $\sim \tau_F$) as produced by, say,
ring oscillators. Such a train of pulses would stabilize the AC signal and
should permit its observation with current technology.  The transient AC regime
is however not restricted to well defined interferometers and could be observed
in any situation where the electronic transport is coherent.  One could consider
using multichannel systems such as small pillars of metallic multilayers (with
metal--metal interfaces acting as the barriers), or the universal conductance
fluctuations of disordered metals. The corresponding signals from the different
channels/set of interfering trajectories would be added incoherently in such
systems which would lead to a sharp peak at the frequency $eV_b/h$ in the
current noise.  In contrast to its superconducting counterpart, the AC frequency
predicted here is not limited by a superconducting gap, meaning that high
frequencies can be obtained. The effect could possibly be used to make tunable
radiofrequency sources that go beyond the GHz regime.

\bigskip
\noindent \tb{Methods}\newline
\noindent\tb{Numerical method.}  The time-dependent numerical
calculations have been performed using the T-Kwant algorithm described
in~\cite{Twave_formalism}, following  the model detailed in~\cite{fabryperot}.
The DC calculations were performed with the Kwant open source
package~\cite{Kwant_preparation}. 

\begin{thebibliography}{21}
\expandafter\ifx\csname natexlab\endcsname\relax\def\natexlab#1{#1}\fi
\expandafter\ifx\csname bibnamefont\endcsname\relax
  \def\bibnamefont#1{#1}\fi
\expandafter\ifx\csname bibfnamefont\endcsname\relax
  \def\bibfnamefont#1{#1}\fi
\expandafter\ifx\csname citenamefont\endcsname\relax
  \def\citenamefont#1{#1}\fi
\expandafter\ifx\csname url\endcsname\relax
  \def\url#1{\texttt{#1}}\fi
\expandafter\ifx\csname urlprefix\endcsname\relax\def\urlprefix{URL }\fi
\providecommand{\bibinfo}[2]{#2}
\providecommand{\eprint}[2][]{\url{#2}}

\bibitem[{\citenamefont{Batelaan and Tonomura}(2009)}]{Batelaan2009}
\bibinfo{author}{\bibfnamefont{A.}~\bibnamefont{Batelaan}} \bibnamefont{and}
  \bibinfo{author}{\bibfnamefont{A.}~\bibnamefont{Tonomura}},
  \bibinfo{journal}{Physics Today} \textbf{\bibinfo{volume}{62}},
  \bibinfo{pages}{38} (\bibinfo{year}{2009}).

\bibitem[{\citenamefont{Imry}(2002)}]{Imry2002}
\bibinfo{author}{\bibfnamefont{Y.}~\bibnamefont{Imry}},
  \emph{\bibinfo{title}{Introduction to mesoscopic physics}}
  (\bibinfo{publisher}{Oxford University Press}, \bibinfo{year}{2002}),
  \bibinfo{edition}{2nd} ed.

\bibitem[{\citenamefont{L\'evy et~al.}(1990)\citenamefont{L\'evy, Dolan,
  Dunsmuir, and Bouchiat}}]{Levy1990}
\bibinfo{author}{\bibfnamefont{L.~P.} \bibnamefont{L\'evy}},
  \bibinfo{author}{\bibfnamefont{G.}~\bibnamefont{Dolan}},
  \bibinfo{author}{\bibfnamefont{J.}~\bibnamefont{Dunsmuir}}, \bibnamefont{and}
  \bibinfo{author}{\bibfnamefont{H.}~\bibnamefont{Bouchiat}},
  \bibinfo{journal}{Phys. Rev. Lett.} \textbf{\bibinfo{volume}{64}},
  \bibinfo{pages}{2074} (\bibinfo{year}{1990}).

\bibitem[{\citenamefont{Castellanos-Beltran
  et~al.}(2013)\citenamefont{Castellanos-Beltran, Ngo, Shanks, Jayich, and
  Harris}}]{Harris2013}
\bibinfo{author}{\bibfnamefont{M.~A.} \bibnamefont{Castellanos-Beltran}},
  \bibinfo{author}{\bibfnamefont{D.~Q.} \bibnamefont{Ngo}},
  \bibinfo{author}{\bibfnamefont{W.~E.} \bibnamefont{Shanks}},
  \bibinfo{author}{\bibfnamefont{A.~B.} \bibnamefont{Jayich}},
  \bibnamefont{and} \bibinfo{author}{\bibfnamefont{J.~G.~E.}
  \bibnamefont{Harris}}, \bibinfo{journal}{Phys. Rev. Lett.}
  \textbf{\bibinfo{volume}{110}}, \bibinfo{pages}{156801}
  (\bibinfo{year}{2013}).

\bibitem[{\citenamefont{Josephson}(1962)}]{Josephson1962}
\bibinfo{author}{\bibfnamefont{B.~D.} \bibnamefont{Josephson}},
  \bibinfo{journal}{Physics Letters} \textbf{\bibinfo{volume}{1}},
  \bibinfo{pages}{251 } (\bibinfo{year}{1962}).

\bibitem[{\citenamefont{Parker et~al.}(1967)\citenamefont{Parker, Taylor, and
  Langenberg}}]{Parker1967}
\bibinfo{author}{\bibfnamefont{W.~H.} \bibnamefont{Parker}},
  \bibinfo{author}{\bibfnamefont{B.~N.} \bibnamefont{Taylor}},
  \bibnamefont{and} \bibinfo{author}{\bibfnamefont{D.~N.}
  \bibnamefont{Langenberg}}, \bibinfo{journal}{Phys. Rev. Lett.}
  \textbf{\bibinfo{volume}{18}}, \bibinfo{pages}{287} (\bibinfo{year}{1967}).

\bibitem[{\citenamefont{Devoret and Schoelkopf}(2013)}]{Devoret2013}
\bibinfo{author}{\bibfnamefont{M.~H.} \bibnamefont{Devoret}} \bibnamefont{and}
  \bibinfo{author}{\bibfnamefont{R.~J.} \bibnamefont{Schoelkopf}},
  \bibinfo{journal}{Science} \textbf{\bibinfo{volume}{339}},
  \bibinfo{pages}{1169} (\bibinfo{year}{2013}).

\bibitem[{\citenamefont{Gennes}(1999)}]{degennesbook}
\bibinfo{author}{\bibfnamefont{P.~G.~D.} \bibnamefont{Gennes}},
  \emph{\bibinfo{title}{Superconductivity Of Metals And Alloys}}
  (\bibinfo{publisher}{Westview Press}, \bibinfo{year}{1999}).

\bibitem[{\citenamefont{Ji et~al.}(2003)\citenamefont{Ji, Chung, Sprinzak,
  Heiblum, Mahalu, and Shtrikman}}]{Heiblum_MachZender}
\bibinfo{author}{\bibfnamefont{Y.}~\bibnamefont{Ji}},
  \bibinfo{author}{\bibfnamefont{Y.}~\bibnamefont{Chung}},
  \bibinfo{author}{\bibfnamefont{D.}~\bibnamefont{Sprinzak}},
  \bibinfo{author}{\bibfnamefont{M.}~\bibnamefont{Heiblum}},
  \bibinfo{author}{\bibfnamefont{D.}~\bibnamefont{Mahalu}}, \bibnamefont{and}
  \bibinfo{author}{\bibfnamefont{H.}~\bibnamefont{Shtrikman}},
  \bibinfo{journal}{Nature} \textbf{\bibinfo{volume}{422}},
  \bibinfo{pages}{415} (\bibinfo{year}{2003}).

\bibitem[{\citenamefont{Roulleau et~al.}(2008)\citenamefont{Roulleau, Portier,
  Roche, Cavanna, Faini, Gennser, and Mailly}}]{20mum_20mK}
\bibinfo{author}{\bibfnamefont{P.}~\bibnamefont{Roulleau}},
  \bibinfo{author}{\bibfnamefont{F.}~\bibnamefont{Portier}},
  \bibinfo{author}{\bibfnamefont{P.}~\bibnamefont{Roche}},
  \bibinfo{author}{\bibfnamefont{A.}~\bibnamefont{Cavanna}},
  \bibinfo{author}{\bibfnamefont{G.}~\bibnamefont{Faini}},
  \bibinfo{author}{\bibfnamefont{U.}~\bibnamefont{Gennser}}, \bibnamefont{and}
  \bibinfo{author}{\bibfnamefont{D.}~\bibnamefont{Mailly}},
  \bibinfo{journal}{Phys. Rev. Lett.} \textbf{\bibinfo{volume}{100}},
  \bibinfo{pages}{126802} (\bibinfo{year}{2008}).

\bibitem[{\citenamefont{Gaury et~al.}(2014)\citenamefont{Gaury, Weston, Santin,
  Houzet, Groth, and Waintal}}]{Twave_formalism}
\bibinfo{author}{\bibfnamefont{B.}~\bibnamefont{Gaury}},
  \bibinfo{author}{\bibfnamefont{J.}~\bibnamefont{Weston}},
  \bibinfo{author}{\bibfnamefont{M.}~\bibnamefont{Santin}},
  \bibinfo{author}{\bibfnamefont{M.}~\bibnamefont{Houzet}},
  \bibinfo{author}{\bibfnamefont{C.}~\bibnamefont{Groth}}, \bibnamefont{and}
  \bibinfo{author}{\bibfnamefont{X.}~\bibnamefont{Waintal}},
  \bibinfo{journal}{Physics Reports} \textbf{\bibinfo{volume}{534}},
  \bibinfo{pages}{1 } (\bibinfo{year}{2014}).

\bibitem[{\citenamefont{Gaury and Waintal}(2014)}]{fabryperot}
\bibinfo{author}{\bibfnamefont{B.}~\bibnamefont{Gaury}} \bibnamefont{and}
  \bibinfo{author}{\bibfnamefont{X.}~\bibnamefont{Waintal}},
  \bibinfo{journal}{Nat. Commun.} \textbf{\bibinfo{volume}{5}}
  (\bibinfo{year}{2014}).

\bibitem[{\citenamefont{Levitov et~al.}(1996)\citenamefont{Levitov, Lee, and
  Lesovik}}]{Levitov1996}
\bibinfo{author}{\bibfnamefont{L.~S.} \bibnamefont{Levitov}},
  \bibinfo{author}{\bibfnamefont{H.}~\bibnamefont{Lee}}, \bibnamefont{and}
  \bibinfo{author}{\bibfnamefont{G.~B.} \bibnamefont{Lesovik}},
  \bibinfo{journal}{Journal of Mathematical Physics}
  \textbf{\bibinfo{volume}{37}}, \bibinfo{pages}{4845} (\bibinfo{year}{1996}).

\bibitem[{\citenamefont{Liang et~al.}(2001)\citenamefont{Liang, Bockrath,
  Bozovic, Hafner, Tinkham, and Park}}]{FabryPerot_carbon_nano}
\bibinfo{author}{\bibfnamefont{W.}~\bibnamefont{Liang}},
  \bibinfo{author}{\bibfnamefont{M.}~\bibnamefont{Bockrath}},
  \bibinfo{author}{\bibfnamefont{D.}~\bibnamefont{Bozovic}},
  \bibinfo{author}{\bibfnamefont{J.~H.} \bibnamefont{Hafner}},
  \bibinfo{author}{\bibfnamefont{M.}~\bibnamefont{Tinkham}}, \bibnamefont{and}
  \bibinfo{author}{\bibfnamefont{H.}~\bibnamefont{Park}},
  \bibinfo{journal}{Nature} \textbf{\bibinfo{volume}{411}},
  \bibinfo{pages}{665} (\bibinfo{year}{2001}).

\bibitem[{\citenamefont{van Wees et~al.}(1989)\citenamefont{van Wees,
  Kouwenhoven, Harmans, Williamson, Timmering, Broekaart, Foxon, and
  Harris}}]{FabryPerot_pioneer}
\bibinfo{author}{\bibfnamefont{B.~J.} \bibnamefont{van Wees}},
  \bibinfo{author}{\bibfnamefont{L.~P.} \bibnamefont{Kouwenhoven}},
  \bibinfo{author}{\bibfnamefont{C.~J. P.~M.} \bibnamefont{Harmans}},
  \bibinfo{author}{\bibfnamefont{J.~G.} \bibnamefont{Williamson}},
  \bibinfo{author}{\bibfnamefont{C.~E.} \bibnamefont{Timmering}},
  \bibinfo{author}{\bibfnamefont{M.~E.~I.} \bibnamefont{Broekaart}},
  \bibinfo{author}{\bibfnamefont{C.~T.} \bibnamefont{Foxon}}, \bibnamefont{and}
  \bibinfo{author}{\bibfnamefont{J.~J.} \bibnamefont{Harris}},
  \bibinfo{journal}{Phys. Rev. Lett.} \textbf{\bibinfo{volume}{62}},
  \bibinfo{pages}{2523} (\bibinfo{year}{1989}).

\bibitem[{\citenamefont{Kretinin et~al.}(2010)\citenamefont{Kretinin,
  Popovitz-Biro, Mahalu, and Shtrikman}}]{InAsFP}
\bibinfo{author}{\bibfnamefont{A.}~\bibnamefont{Kretinin}},
  \bibinfo{author}{\bibfnamefont{R.}~\bibnamefont{Popovitz-Biro}},
  \bibinfo{author}{\bibfnamefont{D.}~\bibnamefont{Mahalu}}, \bibnamefont{and}
  \bibinfo{author}{\bibfnamefont{H.}~\bibnamefont{Shtrikman}},
  \bibinfo{journal}{Nano Letters} \textbf{\bibinfo{volume}{10}},
  \bibinfo{pages}{3439} (\bibinfo{year}{2010}).

\bibitem[{\citenamefont{Beenakker and van Houten}(1991)}]{Beenakker1991}
\bibinfo{author}{\bibfnamefont{C.~W.~J.} \bibnamefont{Beenakker}}
  \bibnamefont{and} \bibinfo{author}{\bibfnamefont{H.}~\bibnamefont{van
  Houten}}, \bibinfo{journal}{Phys. Rev. Lett.} \textbf{\bibinfo{volume}{66}},
  \bibinfo{pages}{3056} (\bibinfo{year}{1991}).

\bibitem[{\citenamefont{Fève et~al.}(2007)\citenamefont{Fève, Mahé, Berroir,
  Kontos, Plaçais, Glattli, Cavanna, Etienne, and Jin}}]{Single_e_source}
\bibinfo{author}{\bibfnamefont{G.}~\bibnamefont{Fève}},
  \bibinfo{author}{\bibfnamefont{A.}~\bibnamefont{Mahé}},
  \bibinfo{author}{\bibfnamefont{J.-M.} \bibnamefont{Berroir}},
  \bibinfo{author}{\bibfnamefont{T.}~\bibnamefont{Kontos}},
  \bibinfo{author}{\bibfnamefont{B.}~\bibnamefont{Plaçais}},
  \bibinfo{author}{\bibfnamefont{D.~C.} \bibnamefont{Glattli}},
  \bibinfo{author}{\bibfnamefont{A.}~\bibnamefont{Cavanna}},
  \bibinfo{author}{\bibfnamefont{B.}~\bibnamefont{Etienne}}, \bibnamefont{and}
  \bibinfo{author}{\bibfnamefont{Y.}~\bibnamefont{Jin}},
  \bibinfo{journal}{Science} \textbf{\bibinfo{volume}{316}},
  \bibinfo{pages}{1169} (\bibinfo{year}{2007}).

\bibitem[{\citenamefont{Dubois et~al.}(2013)\citenamefont{Dubois, Jullien,
  Portier, Roche, Cavanna, Jin, Wegscheider, Roulleau, and
  Glattli}}]{Glattli2013}
\bibinfo{author}{\bibfnamefont{J.}~\bibnamefont{Dubois}},
  \bibinfo{author}{\bibfnamefont{T.}~\bibnamefont{Jullien}},
  \bibinfo{author}{\bibfnamefont{F.}~\bibnamefont{Portier}},
  \bibinfo{author}{\bibfnamefont{P.}~\bibnamefont{Roche}},
  \bibinfo{author}{\bibfnamefont{A.}~\bibnamefont{Cavanna}},
  \bibinfo{author}{\bibfnamefont{Y.}~\bibnamefont{Jin}},
  \bibinfo{author}{\bibfnamefont{W.}~\bibnamefont{Wegscheider}},
  \bibinfo{author}{\bibfnamefont{P.}~\bibnamefont{Roulleau}}, \bibnamefont{and}
  \bibinfo{author}{\bibfnamefont{D.~C.} \bibnamefont{Glattli}},
  \bibinfo{journal}{Nature} \textbf{\bibinfo{volume}{502}}, \bibinfo{pages}{659
  } (\bibinfo{year}{2013}).

\bibitem[{\citenamefont{Zhong et~al.}(2008)\citenamefont{Zhong, Gabor,
  Sharping, Gaetal, and McEuen}}]{McEuen2008}
\bibinfo{author}{\bibfnamefont{Z.}~\bibnamefont{Zhong}},
  \bibinfo{author}{\bibfnamefont{N.~M.} \bibnamefont{Gabor}},
  \bibinfo{author}{\bibfnamefont{J.~E.} \bibnamefont{Sharping}},
  \bibinfo{author}{\bibfnamefont{A.}~\bibnamefont{Gaetal}}, \bibnamefont{and}
  \bibinfo{author}{\bibfnamefont{P.~L.} \bibnamefont{McEuen}},
  \bibinfo{journal}{Nature Nanotechnology} \textbf{\bibinfo{volume}{3}},
  \bibinfo{pages}{201} (\bibinfo{year}{2008}).

\bibitem[{\citenamefont{{Groth} et~al.}(2014)\citenamefont{{Groth}, {Wimmer},
  {Akhmerov}, and {Waintal}}}]{Kwant_preparation}
\bibinfo{author}{\bibfnamefont{C.~W.} \bibnamefont{{Groth}}},
  \bibinfo{author}{\bibfnamefont{M.}~\bibnamefont{{Wimmer}}},
  \bibinfo{author}{\bibfnamefont{A.~R.} \bibnamefont{{Akhmerov}}},
  \bibnamefont{and}
  \bibinfo{author}{\bibfnamefont{X.}~\bibnamefont{{Waintal}}},
  \bibinfo{journal}{New J. Phys.} \textbf{\bibinfo{volume}{16}},
  \bibinfo{pages}{063065} (\bibinfo{year}{2014}).

\end{thebibliography}

\bigskip
\noindent\tb{Acknowledgements} \newline 
This work was supported by the ERC grant MesoQMC from the
European Union.  We thank J.-P. Brison for careful reading of the manuscript and
useful comments.

\bigskip
\noindent\tb{Author contributions}\newline
X.W. initiated the project. B.G. and J.W. performed the numerical simulations.
X.W., B.G. and J.W. performed the analytical calculations, data analysis and wrote the manuscript.

\bigskip
\noindent\tb{Contact information}\newline
Correspondence and requests for materials should be addressed to X.W. (email: xavier.waintal@cea.fr)

\end{document}